\documentclass[letter,11pt]{article}
\pdfoutput=1 

\usepackage{jcappub} 

\usepackage[normalem]{ulem}

\usepackage[utf8]{inputenc}
\usepackage[T1]{fontenc} 
\usepackage{dsfont}
\usepackage{amsmath,amssymb,calc}
\usepackage{color}
\usepackage{amssymb}
\usepackage{graphicx, epsfig, bm}
\usepackage{hyperref}
\usepackage{soul}
 \usepackage{multicol}
\usepackage{changepage}
\usepackage{subfig}

\usepackage{color}
\usepackage{ifthen}

\def\bdm{\begin{displaymath}}
\def\edm{\end{displaymath}}

\def\barray{\begin{array}}
\def\earray{\end{array}}
\def\be{\begin{equation}}
\def\ee{\end{equation}}
\def\ben{\begin{equation} \nonumber}
\def\een{\end{equation}}
\def\ban{\begin{eqnarray*}}
\def\ean{\end{eqnarray*}}
\def\ba{\begin{eqnarray}}
\def\ea{\end{eqnarray}}
\def\eal{\end{align}}
\def\bal{\begin{align}}

\def\({\left(}
\def\){\right)}
\def\[{\left[}
\def\]{\right]}

\def\by{{\bf{y}}}
\def\ba{{a\left(\by\right)}}

\def\bk{{\bf k}}

\def\bx{{\bf x}}

\definecolor{gold}{rgb}{1.0, 0.84, 0.0}
\definecolor{maroon}{rgb}{.25,0,0}
\definecolor{darkorange}{rgb}{1.0, 0.55, 0.0}
\definecolor{corn}{rgb}{0.98, 0.93, 0.36}
\definecolor{bronze}{rgb}{0.8, 0.5, 0.2}
\definecolor{darkgreen}{cmyk}{0.85,0.2,1.00,0.2}

\begin{document}

\title{On the adiabatic subtraction of cosmological perturbations}

\author{Sofia P. Corb\`{a}}
\affiliation{Amherst Center for Fundamental Interactions, Department of Physics, University of Massachusetts, Amherst, MA 01003, U.S.A.}
\author{and Lorenzo Sorbo}



\emailAdd{spcorba@umass.edu}
\emailAdd{sorbo@umass.edu}

\abstract{Adiabatic subtraction is a popular method of renormalization of observables in quantum field theories on a curved spacetime. When applied to the computation of the power spectra of light ($m\ll H$) fields on de Sitter space with flat Friedmann-Lema\^{i}tre-Robertson-Walker slices, the {\em standard prescriptions} of adiabatic subtraction, traceable back to~\cite{Parker:1969au,Parker:1971pt}, lead to results that are significantly different from the standard predictions of inflation not only in the ultraviolet ($k\gg aH$) but also at intermediate ($m\ll k/a\lesssim H$) wavelengths. In this paper we review those results and we contrast them with the power spectra obtained using an alternative prescription for adiabatic subtraction applied to quantum field theoretical systems by Dabrowski and Dunne~\cite{Dabrowski:2014ica,Dabrowski:2016tsx}. This prescription eliminates the intermediate-wavelength effects of renormalization that are found when using the standard one. 
}

\maketitle


\section{Introduction}%

One of the key predictions of primordial inflation is that quantum fluctuations of light fields are amplified into the seeds of the large scale structure of our Universe. As is the case on Minkowskian backgrounds, also for quantum fields on a curved spacetime one needs to address the presence of ultraviolet divergences in the expressions of the physical observables. On Minkowskian backgrounds, divergences of observables that are quadratic in the fields (such as the total energy of a quantum field) can be subtracted ``by hand'', an operation, associated to the normal ordering of creation and annihilation operators, that is justified by the fact that the divergent quantities are unobservable constants. Things get more subtle in time-dependent and/or curved backgrounds.

In this article we study the renormalization of the power spectrum of scalar fluctuations during and after inflation. We will focus on the method of {\em adiabatic subtraction}, the original version of which was proposed by Parker~\cite{Parker:1969au,Parker:1971pt}  to renormalize the energy-momentum tensor of scalar fields in expanding universes. This method leads to finite observables by subtracting from the bare results, mode by mode, the same quantities obtained by replacing the mode functions with their positive--frequency Wentzel-Kramers-Brillouin (WKB) approximation. This is a natural generalization of the subtraction of divergent constants performed to obtain finite results on trivial backgrounds. The WKB approximation is obtained as a recursive series, and, according to the standard prescription, the order at which it should be evaluated is related to the degree of  ultraviolet divergence of the operator that we try to renormalize.  More specifically, one is instructed~\cite{Parker:1969au,Parker:1971pt,Parker:1974qw,Birrell:1982ix,Parker:2009uva} to truncate the WKB series to the minimum order that allows to obtain a UV-finite result.

\begin{figure}[t]
\centering
\includegraphics[scale=.8]{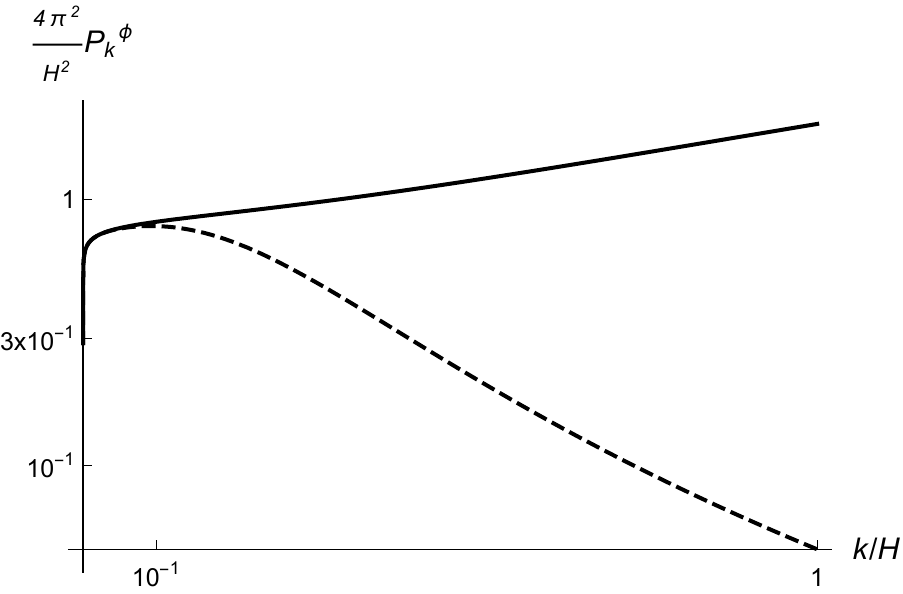} 
\caption{The power spectrum of a scalar field of mass $m^2=.1\,H^2$ on de Sitter space. The solid line represents the non-renormalized result. The dashed line gives the normalized result according to eq.~(18) of~\cite{Parker:2007ni}.}
\label{fig:intro}
\end{figure}

This method has been applied by Parker~\cite{Parker:2007ni} to the standard calculation of the power spectrum ${\cal P}^\phi_k$ of a light ($m\ll H$) scalar field during inflation with Hubble parameter $H$.  In that paper it was found that the renormalized power spectrum converges to the standard result $\frac{H^2}{4\pi^2}$ for $k/a\ll m\ll H$, while for $m\ll k/a\lesssim H$ one can evaluate it to $\sim \frac{H^2}{4\pi^2}\times\frac{3\,m^2\,a^2}{4\,k^2}$. Related analyses can be found in~\cite{Agullo:2008ka,Agullo:2009vq,Glenz:2009zn,Agullo:2009zi}. While the large scale results match the standard expectation, at intermediate scales, where causality arguments would require quasi-constant power spectra, the renormalized spectra show a rapid running. This behavior can be seen for instance in Figure~\ref{fig:intro}, where we show the non-renormalized and the renormalized power spectra (obtained from eq.~(18) of~\cite{Parker:2007ni}) of a scalar with mass $m^2=.1\,H^2$.

Several authors have discussed this result. Shortly after~\cite{Parker:2007ni}, it was   pointed out in~\cite{Finelli:2007fr} that the two-point function, being finite when computed at distinct points, should not need renormalization. The authors of~\cite{Finelli:2007fr}, however, also noted that adiabatic regularization of the power spectrum, a fundamental tool to renormalize the stress-energy tensor, leads to ``unpleasant features'' in the regularized power spectra that persist if one considers the next order in the WKB series.  In~\cite{Urakawa:2009xaa} it has been noted that a time-dependent value of the Hubble parameter makes the adiabatically subtracted component smaller at later times. This view was restated in~\cite{Durrer:2009ii}, where it was stressed that the effect of renormalization should not affect the cosmological scales $k\ll a\,H$.  In~\cite{Marozzi:2011da}, the same authors argued that these problems are alleviated if one assumes a radiation-dominated period prior to inflation, that effectively provides an infrared cut-off to the modes of the scalar perturbations. The paper~\cite{Agullo:2011qg} was written in response to these objections. In \cite{Markkanen:2017rvi} it was shown that adiabatic subtraction can be recast in the form of redefinition of parameters of the Lagrangian. Very recently, finally, the authors of~\cite{Animali:2022lig} have argued that the unusual behavior for superhorizon modes with physical wavelengths shorter than $m^{-1}$ is eliminated by implementing adiabatic subtraction only for modes with wavelength shorter than an infrared cutoff.

The result of~\cite{Parker:2007ni} highlights a couple of undesirable features of the standard formulation of adiabatic subtraction:

\begin{enumerate}

\item adiabatic subtraction leads to what artifacts (significant corrections to the classical result far from the UV regime) for  modes that are not adiabatically evolving. One such example is given by the modes with $m\lesssim k/a\lesssim H$, see Figure~\ref{fig:intro}. One can argue that this is not so upsetting, because if the proper frequencies are not adiabatically evolving, then the concept of particle itself becomes ill-defined. Nevertheless, it would be preferable to {\em (i)} have a trustworthy definition of integral quantities, such as, e.g., $\langle \phi({\bf x},\,t)^2\rangle$, that is given as the integral on all scales of the power spectrum; and {\em (ii)} that the value of the power spectrum takes physically sensible values at all wavelengths;

\item in the standard, textbook~\cite{Birrell:1982ix,Parker:2009uva} prescription for adiabatic subtraction, the order of WKB approximation depends on the degree of ultraviolet  divergence of the operator under consideration. For instance, the calculation of $\langle \phi({\bf x},\,t)^2\rangle$ will require subtraction up to the second order in the WKB expansion, while to compute $\langle \left(\nabla\phi({\bf x},\,t)\right)^2\rangle$ one needs to go to fourth order. A prescription where the order of truncation of the WKB expansion is independent of the operator under consideration might be preferable.

\end{enumerate}

A related subtlety is that adiabaticity depends on the choice of time. For a massless scalar in cosmic time $t$ the proper frequency is given by $\omega_k^2=k^2\,e^{-2Ht}-\frac94 H^2$, that behaves adiabatically, $|d\omega_k/dt|\ll |\omega_k|^2$ at late times $t\to\infty$. On the other hand, in conformal time $\tau$, $d\tau=e^{-Ht}\,dt$, the proper frequency of canonically normalized modes $\omega_k^2=k^2-\frac{2}{\tau^2}$ does {\em not} behave adiabatically at late times, as $|d\omega_k/d\tau|\simeq |\omega_k|^2/\sqrt{2}$ for $\tau\to 0^-$. The notion of adiabaticity thus seems to depend on the choice of time. While this ambiguity can be dealt with by considering terms such as $\frac{\ddot{a}}{a}$ appearing in the proper frequency as higher order terms in the adiabatic expansion, we will show below that this issue is resolved by assuming a definition of time that maintains $\omega_k(t)^2>0$ at all moments. Such a definition of time has been used in the past, see for instance~\cite{Parker:1974qw,Glenz:2009zn}.

In this paper we will explore the implications, for the calculation of power spectra, of an alternative prescription for the order of truncation of the WKB approximation. This prescription, based on the findings in~\cite{dingle,Barry:1989zz,Barry:1990}, has been explicitly applied to quantum field theoretical systems in~\cite{Dabrowski:2014ica,Dabrowski:2016tsx}. As we will discuss in Sections~\ref{sec:bogodef} and~\ref{sec:truncation} below, the adiabatic expansion is generally an asymptotic expansion and as such has an optimal truncation, i.e., there is an order of the expansion that gives an exponentially good approximation to the exact solution. This order has nothing to do with the degree of divergence of the operator one has to renormalize but depends on the parameters of the system. This order is also generally dependent on the momentum $k$. At least in principle, this allows us to avoid the generation of unphysical behavior at intermediate momenta such as that observed when applying the textbook prescription~\cite{Birrell:1982ix,Parker:2009uva}.   Remarkably, the asymptotic behavior of the mode functions which underlies the adiabatic solution is exactly the manifestation of the notion of particle production, that can be evaluated analytically by focusing on the Stokes Phenomenon~\cite{Dumlu:2010ua}. In other words, if the WKB expansion is not asymptotic and can be resummed exactly, then no particle production will occur.

In Section~\ref{sec:applications}, we apply the general results presented in the previous sections to specific cases of inflationary spectra. We find, in agreement with the result found in~\cite{Parker:2007ni}, that the renormalized spectrum for a massless, minimally coupled scalar in exact de Sitter space with flat spatial slices is identically vanishing. For a light, massive field in exact de Sitter space, on the other hand, the application of the prescription~\cite{Dabrowski:2014ica,Dabrowski:2016tsx} gives, unlike the results of~\cite{Parker:2007ni}, a power spectrum that is approximately constant for all scales $k\lesssim aH$. Finally, we apply the general results discussed above to the case of a massless scalar in an FLRW Universe that performs a smooth transition from a quasi-de Sitter to a radiation-dominated stage.

Section~\ref{sec:conclusions}  contains our conclusions and a discussion of these results.

\section{Bogolyubov coefficients, adiabatic subtraction}
\label{sec:bogodef}

Quantum field theory is plagued with divergences. The simplest divergence that is  encountered is that of the expectation value, for a free theory on a Minkowskian background, of quadratic operators such as the energy density. Such a divergence  is cured by subtracting its (formally infinite) ``vacuum contribution'' by hand, or equivalently by replacing the operators under consideration with their normal-ordered version.

In the case in which the field is quantized on a time-dependent background, however, the vacuum of the theory will also be generally evolving, making the concept of ``vacuum contribution'' ambiguous. In this case, finite expectation values for quadratic operators are usually obtained by applying {\em adiabatic subtraction}, i.e., by subtracting, mode by mode, from the expectation value of the operator under consideration, the expectation value of the same quantity evaluated in the adiabatic approximation~\cite{Parker:1969au}. 

We will now review how this prescription can also be phrased in terms of a time-dependent normal ordering - a picture that can be traced back to~\cite{Grib:1969ruc,Grib:1976pw}.

Consider a general system on a time-dependent background described by a quantum scalar field $\hat\phi(\bx,\,t)$ that we quantize as
\begin{align}\label{def:deco}
\hat\phi(\bx,\,t)=\int\frac{d^3\bk}{(2\pi)^{3/2}}\,e^{i\bk\bx}\left[\phi(k,\,t)\,\hat{a}_\bk+\phi(k,\,t)^*\,\hat{a}_{-\bk}^\dagger\right]\equiv\int\frac{d^3\bk}{(2\pi)^{3/2}}\,e^{i\bk\bx}\,\hat\phi(\bk,\,t)\,,
\end{align}
where $\phi(k,\,t)$ satisfies the equation
\begin{align}\label{eq:kg}
\ddot\phi(k,\,t)+\omega_k(t)^2\,\phi(k,\,t)=0\,.
\end{align}
As we will discuss below, in order for the adiabatic subtraction to be well defined, we require $\omega_k(t)^2$ to be positive. We also assume, for the sake of presentation in this section, that $\omega_k(t)\ge 0$ satisfies the adiabaticity conditions $|\dot\omega_k|/\omega_k^2\to 0$ and $|\ddot\omega_k|/ \omega_k^3\to 0$ both as $t\to-\infty$ and as $t\to+\infty$, while at intermediate times $\omega_k(t)^2$ will generally evolve non adiabatically. Under these conditions, the general solution to eq.~(\ref{eq:kg}) for $t\to\pm\infty$ is a linear combination of $\phi^{{\rm ad}}_{(0)}(k,\,t)$ and $\phi^{{\rm ad}}_{(0)}(k,\,t)^*$, with 
\begin{align}
\phi^{{\rm ad}}_{(0)}(k,\,t)=\frac{1}{\sqrt{2\,W_{(0)}(k,\,t)}}e^{-i\int_{t_0}^tW_{(0)}(k,\,t')\,dt'}\,,\qquad W_{(0)}(k,\,t)\equiv \omega_k(t)\ge 0\,,
\end{align}
where the value of $t_0$ (as long as it is real) is irrelevant.

We choose the initial condition to be positive frequency only
\begin{align}
\phi(k,\,t\to-\infty)=\phi^{{\rm ad}}_{(0)}(k,\,t)\,,
\end{align}
which implies that the operator $\hat{a}_\bk\, (\hat{a}_\bk^{\dagger})$ annihilates (creates) quanta of $\phi$ at $t\to -\infty$. 

Then we solve eq.~(\ref{eq:kg}). Since the adiabaticity condition is assumed to be satisfied at late times,  for $t\to +\infty$ the solution must take the form
\begin{align}\label{eq:sollarge0}
\phi(k,\,t\to+\infty)=\alpha(k)\,\phi^{{\rm ad}}_{(0)}(k,\,t)+\beta(k)\,\phi^{{\rm ad}}_{(0)}(k,\,t)^*\,,
\end{align}
where $\alpha(k)$ and $\beta(k)$ are constants: the {\em Bogolyubov coefficients}. Thus, the field $\hat\phi(\bx,\,t)$ at late times reads
\begin{align}
\hat\phi(\bx,\,t\to +\infty)&=\int\frac{d^3\bk}{(2\pi)^{3/2}}e^{i\bk\bx}\left\{\left[\alpha(k)\,\phi^{{\rm ad}}_{(0)}(k,\,t)+\beta(k)\,\phi^{{\rm ad}}_{(0)}(k,\,t)^*\right]\,\hat{a}_\bk\right.\nonumber\\
&\left.\qquad\qquad\qquad\quad+\left[\alpha(k)^*\,\phi^{{\rm ad}}_{(0)}(k,\,t)^*+\beta(k)^*\,\phi^{{\rm ad}}_{(0)}(k,\,t)\right]\,\hat{a}_{-\bk}^\dagger\right\}\,.
\end{align}
We can therefore define the new operators 
\begin{align}\label{eq:new_creat}
&\hat{b}_\bk\equiv\alpha(k)\,\hat{a}_\bk+\beta(k)^*\,\hat{a}_{-\bk}^\dagger\nonumber\\
&\hat{b}^\dagger_\bk\equiv\alpha(k)^*\,\hat{a}^\dagger_\bk+\beta(k)\,\hat{a}_{-\bk}\,,
\end{align}
in terms of which the field $\hat\phi(\bx,\,t)$ takes the form
\begin{align}
\hat\phi(\bx,\,t\to +\infty)=\int\frac{d^3\bk}{(2\pi)^{3/2}}e^{i\bk\bx}\left[\phi^{{\rm ad}}_{(0)}(k,\,t)\,\hat{b}_{\bk}+\phi^{{\rm ad}}_{(0)}(k,\,t)^*\,\hat{b}_{-\bk}^\dagger\,\right]\,.
\end{align}

This equation shows that $\hat{b}_\bk$  is seen as an annihilation operator by an observer born at $t\to+\infty$. In the literature~\cite{Grib:1976pw,Grib:1969ruc} the choice of these creation/annihilation operators is justified by the fact that these are the operators that diagonalize the Hamiltonian at late times.

Let us now evaluate, for instance, the Hamiltonian operator $\hat{H}(t)$, which is time dependent since the system is on a time-dependent background. It is easy to see that
\begin{align}
&\hat{H}(t\to-\infty)=\int d^3\bk\frac{\omega_k(t\to-\infty)}{2}\left(\hat{a}_\bk\,\hat{a}^\dagger_\bk+\,\hat{a}^\dagger_\bk\,\hat{a}_\bk\right)\,,\nonumber\\
&\hat{H}(t\to+\infty)=\int d^3\bk\frac{\omega_k(t\to+\infty)}{2}\left(\hat{b}_\bk\,\hat{b}^\dagger_\bk+\,\hat{b}^\dagger_\bk\,\hat{b}_\bk\right)\,,
\end{align}
whose vacuum expectation value $\langle 0|\hat{H}(t\to\pm\infty)|0\rangle$ is divergent. This divergence can be eliminated by computing the expectation value of the {\em normal ordered} Hamiltonian operator instead, $\langle 0|:\hat{H}:|0\rangle$. {\em We require that an early observer normal orders the $\hat{a}_\bk$ and $\hat{a}^{\dagger}_\bk$ operators, whereas a late observer normal orders the $\hat{b}_\bk$ and $\hat{b}^{\dagger}_\bk$ operators.} On the other hand, since we are working in the Heisenberg picture, the state $|0\rangle$ is independent of time, so that it is annihilated by the $\hat{a}_\bk$ operators, but not by the $\hat{b}_\bk$ operators.

The consequence of this prescription is that 
\begin{align}
\langle 0|:\hat{H}(t\to -\infty):|0\rangle\to 0\,,\qquad \langle 0|:\hat{H}(t\to +\infty):|0\rangle\to \int d^3\bk\,\omega_k(t\to+\infty)\,|\beta(k)|^2\,,
\end{align}
so that, as it is well known, $|\beta(k)|^2$ can be interpreted as the occupation number of created particles at late times.

We can generalize the above procedure to any quantity that is quadratic in the fields. In particular, this can be applied to the power spectrum ${\cal P}_k^\phi$, that we define through
\begin{align}
\langle 0|\hat\phi(\bx,\,t)\hat\phi(\by,\,t)|0\rangle=\int\frac{d^3\bk}{4\pi\,k^3}e^{i\bk(\bx-\by)}\,{\cal P}^\phi_k(t)
\end{align}
which gives ${\cal P}^\phi_k(t)=\frac{k^3}{2\pi^2}\left|\phi(k,\,t)\right|^2$. A straightforward calculation then shows, using eq.~(\ref{eq:sollarge0}), that the normal ordered power spectrum can be written, both for $t\to-\infty$ and for $t\to+\infty$, as
\begin{align}
{\cal P}^\phi_k(t\to\pm\infty)=\frac{k^3}{2\pi^2}\left[\left|\phi(k,\,t\to\pm\infty)\right|^2-\frac{\Im\left\{\phi(k,\,t\to\pm\infty)\,\dot\phi(k,\,t\to\pm\infty)^*\right\}}{W_{(0)}(k,\,t\to\pm\infty)}\right]\,,
\end{align}
which can be shown to be equivalent to
\begin{align}\label{eq:pphi_reno}
{\cal P}^\phi_k(t\to\pm\infty)&=\frac{k^3}{2\pi^2}\,\left[\left|\phi(k,\,t\to\pm\infty)\right|^2-|\phi^{\rm ad}_{(0)}(k,\,t\to\pm\infty)|^2\right]\nonumber\\
&=\frac{k^3}{2\pi^2}\,\left[\left|\phi(k,\,t\to\pm\infty)\right|^2-\frac{1}{2\,\omega_k(t\to\pm\infty)}\right]\,.
\end{align}
The first line of this equation  shows that time-dependent normal ordering amounts to subtracting from the original expression the same one with the mode functions evaluated in the adiabatic approximation. This procedure thus justifies in a natural way {\em adiabatic subtraction} as a method of subtracting the divergent part of the expectation values of operators in time-dependent settings. 

However, in the way in which it is described above, adiabatic subtraction is not always sufficient to subtract {\em all} the UV divergences.  A related issue is that one should require a prescription for the definition of occupation numbers also at finite, even if large, times. To deal with these questions, one modifies the above derivation to include {\em higher orders} in the adiabatic expansion in the definition of $\phi^{\rm ad}(k,\,t)$, and to generalize eq.~(\ref{eq:sollarge0}) to all values of the time.

More specifically, we define the $(2n)-$th order adiabatic approximation $\phi^{{\rm ad}}_{(2n)}(k,\,t)$ to the exact solution of eq.~(\ref{eq:kg}) as 
\begin{align}\label{def:phiad2n}
\phi^{{\rm ad}}_{(2n)}(k,\,t)=\frac{1}{\sqrt{2\,W_{(2n)}(k,\,t)}}e^{-i\int_{t_0}^tW_{(2n)}(k,\,t')\,dt'}
\end{align}
where $W_{(2n)}(k,\,t)$ is found as follows.

We start by inserting into eq.~(\ref{eq:kg}) the Ansatz
\begin{align}
\phi(k,\,t)=\frac{1}{\sqrt{2\,W(k,\,t)}}e^{-i\int_{t_0}^tW(k,\,t')\,dt'}\,,
\end{align}
which implies that $W(k,\,t)$ satisfies the equation
\begin{align}\label{eq:adiab_exp}
W(k,\,t)^2=\omega_k(t)^2+\sqrt{W(k,\,t)}\frac{d^2}{dt^2}\left(\frac{1}{\sqrt{W(k,\,t)}}\right)\,.
\end{align}
The solution to this equation can be found iteratively by expanding it as a series in time derivatives. Then $W_{(2n)}(k,\,t)$ is found as the truncation at the $(2n)$-th order of this derivative (i.e., adiabatic) expansion:
\begin{align}
&W_{(0)}(k,\,t)=\omega_k(t)\,,\nonumber\\
&W_{(2)}(k,\,t)=\omega_k(t)\left[1+\frac{\sqrt{\omega_k(t)}}{2\,\omega_k(t)^2}\frac{d^2}{dt^2}\left(\frac{1}{\sqrt{\omega_k(t)}}\right)\right]\,,\nonumber\\
&W_{(4)}(k,\,t)=\omega_k(t)\left[1+\frac{\sqrt{\omega_k(t)}}{2\,\omega_k(t)^2}\frac{d^2}{dt^2}\left(\frac{1}{\sqrt{\omega_k(t)}}\right)-\frac{\sqrt{\omega_k(t)}}{4\,\omega_k(t)^2}\frac{d^2}{dt^2}\left(\frac{1}{2\omega_k(t)^2}\frac{d^2}{dt^2}\left(\frac{1}{\sqrt{\omega_k(t)}}\right)\right)\right]\,,\nonumber\\
&W_{(6)}(k,\,t)=\ldots \,.
\end{align}
In particular, eq.~(\ref{eq:pphi_reno}) turns out to be valid because, under the assumption of adiabaticity at early and late times, one has $W_{(2n)}(k,\,t\to\pm\infty)\to W_{(0)}(k,\,t\to\pm\infty)=\omega_k(t\to\pm\infty)$.

Then, we define the Bogolyubov coefficients $\alpha_{(2n)}(k,\,t)$ and $\beta_{(2n)}(k,\,t)$, {\em for all values of the time $t$}, through
\begin{align}\label{eq:sollarge}
&\phi(k,\,t)=\alpha_{(2n)}(k,\,t)\,\phi^{{\rm ad}}_{(2n)}(k,\,t)+\beta_{(2n)}(k,\,t)\,\phi^{{\rm ad}}_{(2n)}(k,\,t)^*\,.
\end{align}

{It is important to stress at this point that the definition~(\ref{eq:sollarge}) of Bogolyubov coefficients requires $\phi^{\rm ad}_{(2n)}(k,\,t)$ and $\phi^{\rm ad}_{(2n)}(k,\,t)^*$ to be linearly independent quantities, which implies that $W_{(2n)}(k,\,t)$ should be real. Since $W_{(2n)}(k,\,t)$ is an approximation of $\omega_k(t)$, {\em Bogolyubov coefficients can be consistently defined only at times for which $\omega_k(t)$ is real.}}

In principle one can also study situations where $\omega_k(t)$ transitions from real to imaginary and then again to real values~\cite{Dufaux:2006ee}, but if we want the occupation number to be defined at all times, we will require $\omega_k(t)$ to be real at all times. We will see this in particular in Section~\ref{sec:applications}, where we will adopt a definition of time for which $\omega_k(t)$ is real also for super-horizon modes.

Finally, by requiring $\omega_k(t)^2\ge 0$ one does not incur in  the problem, found in~\cite{Marozzi:2011da}, of having to deal with an imaginary component in the spectrum in eq.~(\ref{eq:pphi_reno})\footnote{{
Trying to be more sophisticated, in the case in which $\omega_k(t)$ is imaginary,  one might look for two linearly independent solutions of eq~(\ref{eq:kg}) with $\omega_k(t)^2<0$, one complex conjugate of the other, to allow for a decomposition like that of eq.~(\ref{def:deco}). One would then find that it is not possible for any of such decompositions to diagonalize the Hamiltonian.}}.

Eq.~(\ref{eq:sollarge}) does not determine $\alpha_{(2n)}(k,\,t)$ and $\beta_{(2n)}(k,\,t)$ uniquely, so that we need a second prescription. Among various options that give equivalent results at $t\to \pm\infty$, we choose that obtained by taking the time derivative of eq.~(\ref{eq:sollarge}) while keeping $\alpha_{(2n)}(k,\,t)$ and 
$\beta_{(2n)}(k,\,t)$ constant:
\begin{align}\label{eq:sollarge1}
\dot\phi(k,\,t)=&\frac{\alpha_{(2n)}(k,\,t)}{\sqrt{2\,W_{(2n)}(k,\,t)}}\left(-iW_{(2n)}(k,\,t)-\frac{\dot{W}_{(2n)}(k,\,t)}{2\,W_{(2n)}(k,\,t)}\right)e^{-i\int_{t_0}^tW_{(2n)}(k,\,t')\,dt'}\nonumber\\
&+\frac{\beta_{(2n)}(k,\,t)}{\sqrt{2\,W_{(2n)}(k,\,t)}}\left(iW_{(2n)}(k,\,t)-\frac{\dot{W}_{(2n)}(k,\,t)}{2\,W_{(2n)}(k,\,t)}\right)e^{i\int_{t_0}^tW_{(2n)}(k,\,t')\,dt'}\,.
\end{align}

Eqs.~(\ref{eq:sollarge}) and~(\ref{eq:sollarge1}) can be inverted to give
\begin{align}
&\alpha_{(2n)}(k,\,t)=\frac12\sqrt{2\,W_{(2n)}(k,\,t)}\,e^{i\int_{t_0}^tW_{(2n)}(k,\,t')\,dt'}\left[\phi(k,\,t)+i\frac{\dot\phi(k,\,t)}{W_{(2n)}(k,\,t)}\right]\,,\nonumber\\
&\beta_{(2n)}(k,\,t)=\frac12\sqrt{2\,W_{(2n)}(k,\,t)}\,e^{-i\int_{t_0}^tW_{(2n)}(k,\,t')\,dt'}\left[\phi(k,\,t)-i\frac{\dot\phi(k,\,t)}{W_{(2n)}(k,\,t)}\right]\,,
\end{align}
and we can thus define the time dependent creation/annihilation operators 
\begin{align}\label{eq:new_creat}
&\hat{b}^{(2n)}_\bk(t)\equiv\alpha_{(2n)}(k,\,t)\,\hat{a}_\bk+\beta_{(2n)}(k,\,t)^*\,\hat{a}_{-\bk}^\dagger\nonumber\\
&\hat{b}^{(2n)}_\bk(t)^\dagger\equiv\alpha_{(2n)}(k,\,t)^*\,\hat{a}^\dagger_\bk+\beta_{(2n)}(k,\,t)\,\hat{a}_{-\bk}\,.
\end{align}
that are seen as annihilation/creation operators for an observer born at time $t$.

In particular, after normal ordering of the $\hat{b}^{(2n)}_\bk(t)$ and $\hat{b}^{(2n)}_\bk(t)^{\dagger}$ operators, the power spectrum of the field $\hat\phi(\bx,\,t)$ reads
\begin{align}\label{eq:powersp_trunc}
{\cal P}_k^\phi(t)=\frac{k^3}{2\pi^2}\left[\left|\phi(k,\,t)\right|^2-\frac{1}{2\,W_{(2n)}(k,\,t)}\right]
\end{align}
where all the dependence on the order $(2n)$ of the adiabatic expansion lies in the $W_{(2n)}(k,\,t)$ at the denominator in the last term, since $\phi(k,\,t)$ is the exact solution to the equation of motion. 

\section{Truncation vs resummation of the adiabatic expansion}
\label{sec:truncation}

As discussed in the previous section, in order to apply adiabatic subtraction we need to truncate the adiabatic expansion at the right adiabatic order. The next questions is then: which order? According to~\cite{Parker:1974qw,Birrell:1982ix,Parker:2009uva}, the prescription is to keep in $W_{(2n)}(k,\,t)$ all the terms, {\em but no more}, that are necessary to cancel all the UV divergences in the operator under consideration.\footnote{{Moreover, the textbook prescription is to further expand $\frac{1}{2\,W_{(2n)}(k,\,t)}$ in eq.~(\ref{eq:powersp_trunc}) in a derivative expansion:
\begin{align}
\frac{1}{2\,W_{(2n)}}\simeq \frac{1}{2\,W_{(0)}}-\frac{W_{(2)}-W_{(0)}}{2\,W_{(0)}^2}+\dots \,\,\,.
\end{align}}} 
 While one might argue that this prescription has the advantage of being the least intrusive way of generating UV-finite results, there are, as we have discussed in the Introduction, a couple of reasons for concern. First, it is not clear why the order of truncation should depend on the observable.  For instance, for the same field, this prescription instructs to use $W_{(2)}(k,\,t)$ when computing the power spectrum, and $W_{(4)}(k,\,t)$ when computing the energy density. Second, while this prescription creates a good behavior for the observables at $k\to\infty$, it often generates artifacts (i.e. features in the observables where the subtracted component overwhelms the bare one) at finite momenta, such as those observed in~\cite{Parker:2007ni}.

In this article we discuss an alternative approach, see~\cite{Dabrowski:2014ica,Dabrowski:2016tsx}, where the order of truncation is not related to the degree of divergence of the operator we try to renormalize. In fact, the WKB series is in general an asymptotic expansion and as such it has an optimal truncation which gives the best possible approximation to the exact solution. Truncating the adiabatic expansion to the optimal order, besides being a less arbitrary choice, also removes infrared artifacts, as the order of the truncation generally depends on the wavenumber $k$ and therefore the subtracted function will not be the same in the UV and the IR regime. More importantly, as we will see below, optimal truncation of the WKB series, at least in the regime of large truncation order, leads to a {\em universal} functional dependence of the Bogolyubov coefficients.

The asymptotic nature of the WKB expansion is associated to the fact that the adiabatic solutions are defined only locally, as we will now discuss.

\begin{figure}[t]
\centering
\includegraphics[scale=.17]{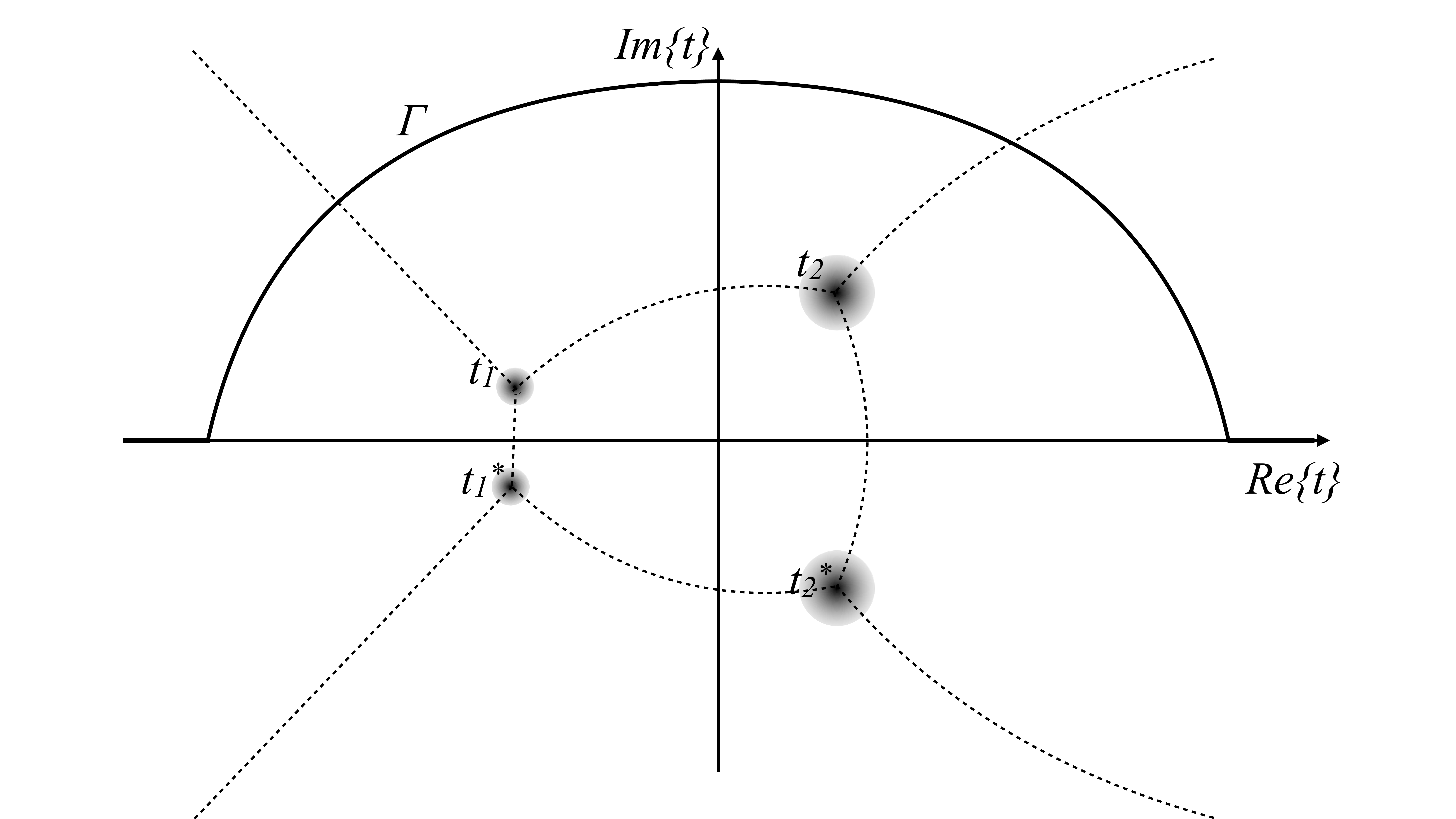} 
\caption{A schematic representation of the complex-$t$ plane. In this example the theory has four turning points, coming in complex conjugate pairs, $t_1$, $t_1^*$, $t_2$, $t_2^*$. Each turning point is surrounded by a shaded area where the adiabaticity condition is violated. The path $\Gamma$ allows us to go from $t\to -\infty$ to $t\to+\infty$ while traveling arbitrarily far from the turning points. However, along its path, it must cross Stokes lines (marked as dotted lines), where a negative frequency component is added to the WKB solution.}
\label{fig:complext}
\end{figure}

The WKB approximation requires the adiabatic conditions $|\dot\omega_k|\ll \omega_k^2$ and $|\ddot\omega_k|\ll \omega_k^3$ to be satisfied. The points in the complex-$t$ plane where $\omega_k=0$ are called {\em poles} or {\em turning points}, and each of them is surrounded by a region where the adiabatic approximation is violated. For the sake of the presentation, let us assume that the adiabatic conditions are satisfied as $t\to \pm \infty$, i.e., that the concept of particle is well defined at early and at late times. Let us also assume that there is a path $\Gamma$ in the complex-$t$ plane, see Figure~\ref{fig:complext}, that allows to go from $t\to-\infty$ to $t\to+\infty$ while staying arbitrarily far from the turning points (which is guaranteed as long as the number of turning points is finite).  This would imply that, starting at $t\to-\infty$ with a positive frequency solution $\propto e^{-i\int^t\omega_k\,dt'}$, and using the WKB approxiation along $\Gamma$, we get solutions that are only positive frequency also as $t\to +\infty$, i.e., $\beta(k,\,t\to+\infty)=0$. Clearly there is something wrong here, since in general, if the adiabatic condition is violated at finite times, particles should be created and  $\beta(k,\,t\to+\infty)$ should not vanish! 

What has gone wrong? The issue is that the WKB approximation is defined only {\em locally}, and the mode functions at initial time cannot be analytically continued  through the whole complex plane. The borders of regions of validity of such local approximations are called {\em Stokes lines}, and the generation of a negative frequency component as one crosses a Stokes line is known as the {\em Stokes phenomenon}~\cite{Stokes:1902}.

The origin of the Stokes phenomenon is that in any given Stokes region the WKB perturbative expansion fails to capture the full expression of the solution, i.e.,  a part of the solution does not appear {\em at any order} in the WKB approximation\footnote{This is similar to the case of a function such as $f(x)=g(x)+e^{-1/x^2}$ where $g(x)$ can be expanded as a Taylor series around $x=0$. If we try to evaluate $f(x)$ at finite values of $x$ by using its Taylor expansion around $x=0$, we will never recover the component $e^{-1/x^2}$, irrespective to the order at which we perform the Taylor expansion.}, and as a consequence is not generated by the simple analytical continuation to $t\to+\infty$ of the positive frequency solution. In other words, the Stokes phenomenon signals the fact that the WKB expansion cannot be resummed everywhere to get the exact mode function, i.e., that the WKB expansion is asymptotic. We thus obtain the equivalence
\begin{align}
{\rm asymptotic\ WKB\ expansion\ }\Leftrightarrow {\rm Stokes\ lines}\Leftrightarrow {\rm frequency\ mixing} \Leftrightarrow {\rm particle\ creation}.\nonumber
\end{align}

On the other hand, if the WKB expansion {\em can} be resummed, then the exact mode functions can be written, in the entire complex $t$ plane, as $\frac{1}{{\sqrt{2W(k,\,t)}}}{e^{-i\int^tW(k,\,t')\,dt'}}$ for some function $W(k,\,t)$ (with boundary condition $W(k,\,t\to-\infty)\to \omega_k(t\to-\infty)$) and there is no particle creation, $\beta(k,\,t)=0$. In this case it is apparent how the prescription~\cite{Dabrowski:2014ica,Dabrowski:2016tsx} is radically different from that of~\cite{Parker:1974qw,Birrell:1982ix,Parker:2009uva}: while~\cite{Dabrowski:2014ica,Dabrowski:2016tsx} leads to $\beta(k,\,t)=0$ if the WKB series can be resummed exactly,~\cite{Parker:1974qw,Birrell:1982ix,Parker:2009uva} would truncate the WKB series to a finite order, yielding $\beta(k,\,t\to+\infty)\neq 0$.

\smallskip

As we have seen above, the Stokes lines represent the boundaries of the regions with well defined local WKB expansions. On those  lines we have the greatest disparity between the exponentials appearing in the positive and negative frequency solutions. Thus, the Stokes lines are determined by the condition
\begin{align}\label{def:singulant}
F_k(t)\equiv-2i\int_{t_c}^t\omega_k(t')\,dt'= \text{purely real}\,,
\end{align}
where we denoted as $t_c$ a turning point ($\omega_k(t_c)= 0$) in the complex-$t$ plane with positive imaginary part. $F_k(t)$ is called the {\em singulant} variable.

Stokes~\cite{Stokes:1902} provided a {\em connection formula} which allows us to find the mode function on one side of the Stokes line once we have it on the other side: in order to have the concordance between different asymptotic representations defined in different regions of the complex $t$ plane, the multiplier of the sub-dominant exponential must have a jump  which is equal to $i$ (the imaginary unit) times the multiplier of the dominant one. If we assume the mode function at early times to be 
\begin{equation}
\phi(k,\,t\ {\text {on one side of Stokes line}})= \frac{1}{\sqrt{2\omega_k(t)}} e^{-i\int_{t_0}^t \omega_k(t')dt'}\,,
\end{equation}
then, after crossing the Stokes line, it will have the form
\begin{equation}
\phi(k,\,t\ {\text {on other side of Stokes line}})= \frac{1}{\sqrt{2\omega_k(t)}} \left[e^{-i\int_{t_0}^t \omega_k(t')dt'}-i\, e^{-2i\int_{t_0}^{t_c} \omega_k(t')dt'} e^{i\int_{t_0}^t \omega_k(t')dt'}\right]\,.
\end{equation}

A  study of the microscopic structure of the Stokes phenomenon has been performed by Dingle~\cite{dingle} and Berry~\cite{Barry:1989zz,Barry:1990}.  These authors have found a formula for the order $n^{\rm optimal}$ at which the WKB expansion should be truncated to yield the best approximation to $\beta(k,\,t)$. More importantly, they have resolved the thickness of the Stokes line,  arguing that the evolution of the multiplier of the subdominant mode function (i.e., the Bogolyubov coefficient $\beta(k,\,t)$) in a neighborhood of the Stokes line has a universal and smooth form, at least as long as $n^{\rm optimal}\gg 1$.   Dunne and Dabrowsky~\cite{Dabrowski:2014ica,Dabrowski:2016tsx} have verified the consistency of those formulae for the cases of Schwinger effect and of creation of massive scalars in a closed de Sitter Universe.

The fact that the optimal truncation of the WKB series gives the best approximation to the universal form of $\beta(k,\,t)$ is the main reason behind the proposal for adiabatic subtraction, alternative to that of~\cite{Birrell:1982ix,Parker:2009uva}, that we will study in this paper.

Such a universal form is obtained in terms of a natural Stokes-line crossing variable $\sigma_k(t)$ given by
\begin{equation}\label{def:sigmakt}
\sigma_k(t)\equiv \frac{\Im\{F_k(t)\}}{\sqrt{2\,\Re\{F_k(t)\}}}\,,
\end{equation}
where the singulant function $F_k(t)$ is defined in eq.~(\ref{def:singulant}). The Bogolyubov coefficient $\beta(k,\,t)$, as we cross the Stokes' line, is given by
\begin{align}\label{eq:betaBerry}
\beta(k,\,t)\simeq \frac{i}{2}\left[1+{\rm erf}\left(\sigma_k(t)\right)\right]\,e^{-F_k^{(0)}}\,,
\end{align}
that indeed ranges from $0$ to $i\,e^{-F_k^{(0)}}$ as we cross the Stokes' line, where
\begin{align}
 F_k^{(0)}= -i\int_{t_c}^{t_c^*}\omega_k(t')\,dt'
\end{align}
is the integral of $\omega_k(t)$ taken along the Stokes' line between two complex conjugate turning points, and is therefore real (and, with appropriate choice of branches, positive).

Also, the order of the optimal truncation of the WKB series is given by
\begin{align}\label{eq:noptimal}
n^{\rm optimal}\simeq {\rm Int}\left[F_k^{(0)}/2\right]
\end{align}
which shows that this whole discussion is strictly speaking valid only as long as $F_k^{(0)}\gg 1$. In fact, the thickness of the Stokes' line, as given by the size of the region across which $\beta(k,\,t)$ accumulates most of its variation, is given by the range of $t$ for which $\sigma_k(t)$ ranges from $O(-1)$ to $O(+1)$, which is shown by eq.~(\ref{def:sigmakt}) to scale as $1/\sqrt{F_k^{(0)}}$. Equivalently, one can estimate the size of the regions about $t_c$ for which the adiabaticity condition is not satisfied. To do so, we assume that $t_c$ is a simple zero {and set without loss of generality  $\Re\{t_c\}=0$.} Then, linearizing, $\omega_k(t)\simeq \dot\omega_k(t_c)\times\left(t-t_c\right)$, we obtain that $|\dot\omega_k(t)| \gtrsim \omega_k(t)^2$ for $|t-t_c|\lesssim 1/\sqrt{|\dot\omega_k(t_c)|}$, so that if we want $t_c$ and $t_c^*$ to be distant enough that their regions of non-adiabaticity do not overlap, we must require $|t_c-t_c^*|=O(|t_c|)\gg 1/\sqrt{|\dot\omega_k(t_c)|}$. Estimating the singulant for the linearized expression of $\omega_k(t)$ we obtain $F_k(t)\approx \int_{t_c}^t\dot\omega_k(t_c)\times\left(t'-t_c\right)\,dt'\approx \dot\omega_k(t_c)\left(t-t_c\right)^2\Rightarrow F_k^{(0)}\approx\dot\omega_k(t_c)\,t_c^2$, so that the condition that the poles are distinct, $O(|t_c|)\gg 1/\sqrt{|\dot\omega_k(t_c)|}$ is equivalent to $F_k^{(0)}\gg 1$.

If the condition $F_k^{(0)}\gg 1$ is not satisfied, then the use of the  expressions discussed above is not justified. In the absence of any specific formula for the case $F_k^{(0)}=O(1)$, however, we will still truncate the WKB series at the value of $n$ for which the first local minimum of $|W_{(2n)}-W_{(2n-2)}|$ is reached.

\section{Adiabatic subtraction for a scalar field in the inflationary Universe}%
\label{sec:applications}

After having seen general prescriptions for adiabatic subtraction, our goal is to apply these results to the case of primordial inflation.

The action of a massive test scalar field $\phi$ on a spatially flat FLRW background is given by
\begin{align}
{\cal S}_\phi=\int \frac{d^3k}{(2\pi)^3}\,dt\,a^3\left(\frac12|\dot\phi(k,\,t)|^2-\frac{k^2}{2\,a^2}\left|\phi(k,\,t)\right|^2-\frac{m^2}{2}\left|\phi(k,\,t)\right|^2\right)
\end{align}

In order to  work with real frequencies at all times, we define a new time variable (this definition of time has been used in the past, see for instance~\cite{Parker:1974qw,Glenz:2009zn})
\begin{align}
d\theta=\frac{dt}{a(t)^3}\,,
\end{align}
such that the action for $\phi(k,\,t)$ reads
\begin{align}
{\cal S}_\phi=\int \frac{d^3k}{(2\pi)^3}\,d\theta\left(\frac12\left|\frac{d\phi(k,\,\theta)}{d\theta}\right|^2-\frac12\left(k^2\,a^4+m^2\,a^6\right)\left|\phi(k,\,\theta)\right|^2\right)\,.
\end{align}
The field $\phi(k,\,\theta)$ is thus already canonically normalized, and the frequency of the mode with momentum $k$ is given by $\omega_k=\sqrt{k^2\,a^4+m^2\,a^6}$, that is positive definite. 

Let us now consider different regimes, starting from the simplest case of a massless field on an exact de Sitter space in flat slicing.

\subsection{A massless scalar on exact de Sitter space in flat slicing}
\label{subsec:massless}

In this case $\omega_k=k\,a^2$ with $a(t)=e^{Ht}$, or, using
\begin{align}
&d\theta=\frac{dt}{a(t)^3}\Rightarrow \theta=-\frac{e^{-3Ht}}{3\,H}\,,\qquad -\infty<\theta<0\nonumber\\
& a(\theta)=\left(-3H\theta\right)^{-1/3}\,,\qquad \omega_k(\theta)=\frac{k}{\left(-3H\theta\right)^{2/3}}
\end{align}
where we have set $\theta=-1/3H$ at the end of inflation, $t=0$, $a=1$.

The equation of motion for mode functions,
\begin{align}
\frac{d^2\phi(k,\,\theta)}{d\theta^2}+\frac{k^2}{\left(-3H\theta\right)^{4/3}}\,\phi(k,\,\theta)=0\,,
\end{align}
with positive frequency only solutions as $\theta\to -\infty$, reads
\begin{align}\label{eq:phik_dSmassless}
\phi(k,\,\theta)&=\frac{H}{\sqrt{2k^3}}\left(1-i\frac{k}{H}\left(-3H\theta\right)^{1/3}\right)e^{i\frac{k}{H}\left(-3H\theta\right)^{1/3}}\nonumber\\
&=\frac{H}{\sqrt{2k^3}}\left(1-i\frac{k}{a\,H}\right)e^{i\frac{k}{a\,H}}\,,
\end{align}
which is the well-known expression for the mode functions of a massless scalar on a de Sitter space with flat slices.

Remarkably, this solution can be written in the form
\begin{align}\label{eq:exact_dSmassless}
\phi(k,\,\theta)=\frac{e^{-i\int^\theta W(k,\,\theta')\,d\theta'}}{\sqrt{2W(k,\,\theta)}}\,,
\end{align}
if we choose
\begin{align}\label{eq:wresumm0}
W(k,\,\theta)=\frac{k^3\,a(\theta)^2}{k^2+H^2\,a(\theta)^2}\,.
\end{align}

Since, in the case of a massless scalar on de Sitter space, the mode functions can be written in the form~(\ref{eq:exact_dSmassless}), the WKB series can be resummed exactly, which implies that the Bogolyubov coefficient $\beta(k,\,t)$ vanishes. One could reach an analogous conclusion by observing that by moving along the real negative $\theta$-axis, we encounter no Stokes lines in the complex-$\theta$ plane, as $\omega(\theta)$ has no zeros at finite $\theta$.

We thus conclude that the prescription we are examining in this paper implies {\em no particle production} for a massless scalar in exact de Sitter space in flat slicing.

Let us compare our result with that of the prescription of~\cite{Birrell:1982ix,Parker:2009uva}. In this case, one is instructed to truncate the WKB series to second order, 
\begin{align}
W_{(2)}(k,\,\theta)=k\,a(\theta)^2\left(1-\frac{H^2\,a(\theta)^2}{k^2}\right)\,,
\end{align}
so that applying eq.~(\ref{eq:powersp_trunc}) with $n=1$ and using eq.~(\ref{eq:phik_dSmassless}) we obtain
\begin{align}\label{eq:pphi_parkermassless}
{\cal P}_k^\phi(\theta)=\frac{H^2}{4\pi^2}\left[\left(1+\frac{k^2}{a(\theta)^2\,H^2}\right)-\left(\frac{k^2}{a(\theta)^2\,H^2}\,\frac{1}{1-\frac{a(\theta)^2\,H^2}{k^2}}\right)\right]
\end{align}
where the term in the first $(...)$ is proportional to $|\phi_k(\theta)|^2$ whereas the second term is proportional to $1/W_{(2n)}(k,\,\theta)$. If, in the spirit of the adiabatic approximation, we expand the second term to $O(k^0)$, we obtain that the prescription~\cite{Parker:1974qw,Birrell:1982ix,Parker:2009uva} leads to ${\cal P}_k^\phi(\theta)=0$, which agrees with the result obtained by setting $m=0$ in~\cite{Parker:2007ni} and, more importantly, which is the same result we obtained above using the prescription of~\cite{Dabrowski:2014ica,Dabrowski:2016tsx}. 

{The vanishing of the power spectrum of a massless, minimally coupled scalar on de Sitter space with flat slices deserves a couple of comments. First off, it is in agreement with results from earlier literature. For instance, the formulae in~\cite{Habib:1999cs}, when applied to this case, would also give a vanishing spectrum. Second, its disagreement with the usual expectation $\langle \phi(\bx,\,t)\rangle\propto H^3\,t$ can be traced back to the absence of an infrared cutoff in de Sitter space {\em in flat slicing}. The usual result $\langle \phi(\bx,\,t)\rangle\propto H^3\,t$ can be indeed recovered in closed de Sitter space, that has a natural infrared cutoff given by the wavelength of the modes that left the horizon when the scale factor of the Universe was at its minimum~\cite{Allen:1987tz,Habib:1999cs}.

Finally, let us note that, even if they end up giving the same result, the (vanishing) power spectra obtained by using the two prescriptions are qualitatively different. In the case of the standard prescription,  the final result is obtained from a truncation followed by a Taylor expansion of the series in $aH/k$, which is therefore reliable only in the limit $k\gg aH$. In the second case the vanishing renormalized spectrum is the result of a resummation of the entire series. As we will see in the next example, the two prescriptions will give, in general, different results.

\subsection{A massive scalar on exact de Sitter space in flat slicing}
 
In the case $m\neq 0$, with a de Sitter background, one can still find an exact solution to the mode equation,
\begin{align}
\phi(k,\,\theta)=\sqrt{\frac{\pi}{4\,H\,a(\theta)^3}}\,e^{i\nu\pi/2+i\pi/4}\,H^{(1)}_\nu\left(\frac{k}{a(\theta)\,H}\right)\,,\qquad \nu\equiv\sqrt{\frac94-\frac{m^2}{H^2}}\,,\quad m<\frac32H
\end{align}
which, however, cannot be written in the form~(\ref{eq:exact_dSmassless}) for any function $W(k,\,\theta)$, as we will now discuss.

First, unlike the massless case, for $m\neq 0$ the frequency $\omega_k=\sqrt{k^2\,a(\theta)^4+m^2\,a(\theta)^6}$ does have zeros at finite values of $\theta$ given by  
\begin{align}
\theta_\pm=\pm\frac{i}{3H}\frac{m^3}{k^3}\,,
\end{align}
from which Stokes lines emanate. One of those lines crosses the real $\theta$ axis at a point $\theta_{\rm Stokes,\,Real}$ determined by solving the equation
\begin{align}
i\int_{\theta_+}^{\theta_{\rm Stokes,\,Real}}\sqrt{k^2\,a(\theta)^4+m^2\,a(\theta)^6}\,d\theta={\rm real}\,,
\end{align}
i.e., changing variable back to physical time $t$ and integrating,
\begin{align}
i\frac{m}{H}\left[\log\left(e^y+\sqrt{e^{2y}+1}\right)-\sqrt{1+e^{-2y}}\right]_{y=Ht_{\rm Stokes,\,Real}-\log(k/m)}-\frac{\pi\,m}{2\,H}={\rm Real}\,,
\end{align}
so that $t_{\rm Stokes,\,Real}$ can be found by solving numerically the equation $\log\left(e^y+\sqrt{e^{2y}+1}\right)-\sqrt{1+e^{-2y}}=0$, yielding
\begin{align}
Ht_{\rm Stokes,\,Real}\simeq .411+\log(k/m)\,.
\end{align}
As a consequence, particle creation happens approximately when the scale factor $e^{Ht}$ crosses $k/m$, a consequence of the de Sitter symmetry $t\to t+\Delta t$, $k\to k\, e^{-H\,\Delta t}$.

The existence of a Stokes line shows that the WKB series is asymptotic. The singulant reads
\begin{align}
F_k^{(0)}=i\int_{\theta_-}^{\theta_+}\sqrt{k^2\,a(\theta)^4+m^2\,a(\theta)^6}\,d\theta=i\int_{[\log(k/m)+i\pi/2]/H}^{[\log(k/m)-i\pi/2]/H}\sqrt{k^2\,e^{-2Ht}+m^2}\,{dt}=\pi\frac{m}{H}\,,
\end{align}
which means that the WKB approximation will be a good one for $m\gg H$. In this case, eq.~(\ref{eq:betaBerry}) gives $|\beta(k,\,t)|^2\propto e^{-2\pi m/H}$ in agreement with derivations of the rate of creation of heavy particles based on Schwarzschild-de Sitter metric, such as that in~\cite{Gibbons:1977mu}.

A second, more pedestrian way to see that the series is asymptotic is to solve the eq.~(\ref{eq:adiab_exp}) by brute force, at least for the first few terms. By using an algebraic manipulation program we obtain
\begin{align}
W(k,\,t)&\sim ka^2\left[\left(1+\frac12\frac{m^2\,a^2}{k^2}+O(m^4)\right)+\frac{a^2\,H^2}{k^2}\left(-1-\frac14\frac{m^2\,a^2}{k^2}+O(m^4)\right)\right.\nonumber\\
&\left.+\frac{a^4\,H^4}{k^4}\left(1-\frac52\frac{m^2\,a^2}{k^2}+O(m^4)\right)+\frac{a^6\,H^6}{k^6}\left(-1+\frac{217}{8}\frac{m^2\,a^2}{k^2}+O(m^4)\right)\right.\nonumber\\
&\left.+\frac{a^8\,H^8}{k^8}\left(1-\frac{3249}{8}\frac{m^2\,a^2}{k^2}+O(m^4)\right)+\frac{a^{10}\,H^{10}}{k^{10}}\left(-1+\frac{39523}{4}\frac{m^2\,a^2}{k^2}+O(m^4)\right)+...\right]\,,
\end{align}
which shows that, while the $O(m^0)$ terms can be resummed to give the massless result~(\ref{eq:wresumm0}), the coefficients of the $O(m^2)$ terms are rapidly increasing, signaling the asymptotic nature of the WKB series in the massive case.

The case $m\ll H$ is the one of greatest phenomenological interest. In this case, the formulae of~\cite{Barry:1989zz,Barry:1990} are not strictly speaking valid, as we discussed at the end of Section~\ref{sec:truncation} above, and we will simply truncate the WKB series where $|W_{(2n+2)}-W_{(2n)}|$ displays a minimum.  In particular, we have
\begin{align}
&W_{(0)}(k,\,t)=a(t)^3\sqrt{q(t)^2+m^2}\,,\,\qquad\qquad \nonumber\\
&W_{(2)}(k,\,t)-W_{(0)}(k,\,t)=-a(t)^3\,H^2\,\frac{9\,m^4+22\,m^2\,q(t)^2+8\,q(t)^4}{8\,\left(q(t)^2+m^2\right)^{5/2}}\,,\qquad q(t)\equiv\frac{k}{a(t)}
\end{align}
so that both for $q(t)\ll m\ll H$ and for $m\ll q(t)\lesssim H$ we have $|W_{(0)}(k,\,t)|\ll |W_{(2)}(k,\,t)-W_{(0)}(k,\,t)|$. As a consequence, for $m\ll H$ and $k\lesssim a\,H$ we we will only keep the zeroth order of the WKB series. This gives the power spectrum
\begin{align}
{\cal P}_k^\phi=\frac{k^3}{2\pi^2}\left[\frac{\pi}{4Ha^3}\left|H^{(1)}_\nu\left(\frac{k}{aH}\right)\right|^2-\frac{1}{2\sqrt{k^2a^4+m^2a^6}}\right]\,,\qquad m\ll H\,,\quad k\lesssim aH\,.
\end{align}
%

\begin{figure}[t]
	\centering
	\includegraphics[scale=.6]{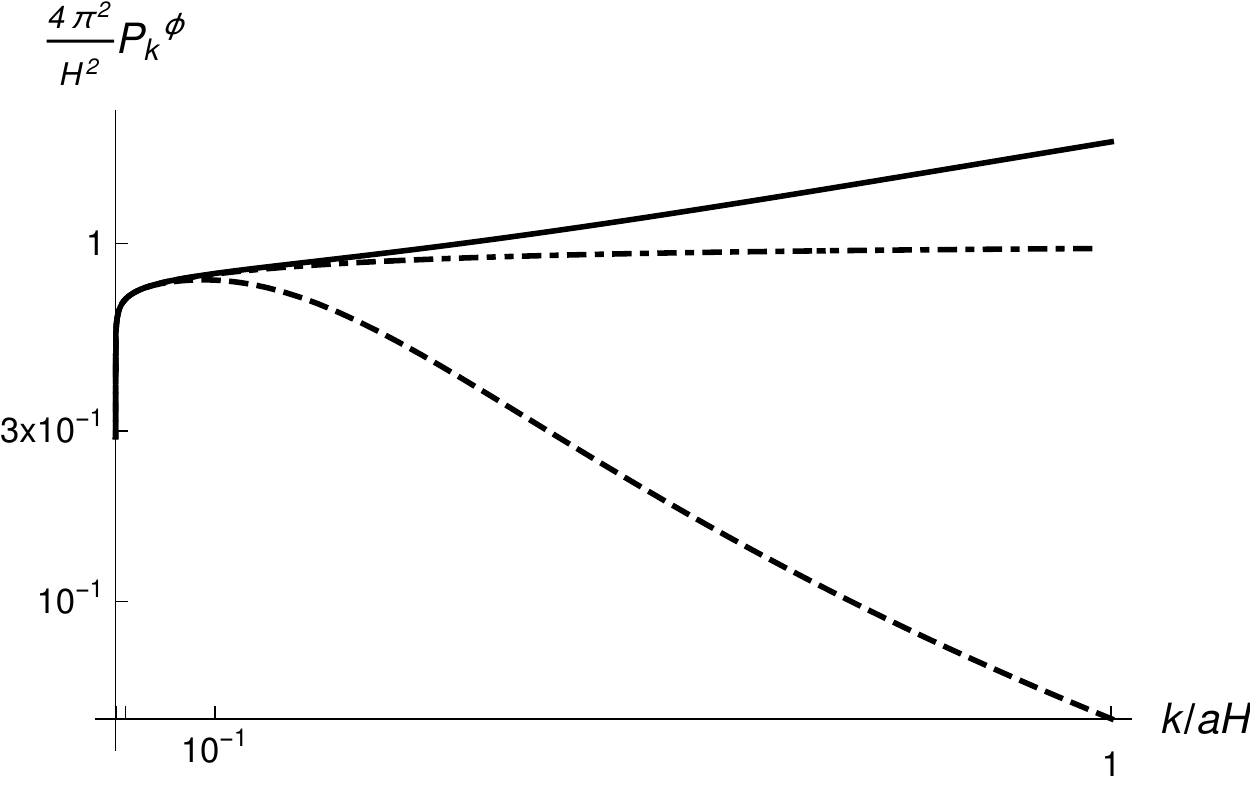} 
	\caption{The power spectrum of a scalar field of mass $m^2=.1H^2$ on de Sitter space. The solid line represents the non-renormalized result. The dashed line shows the normalized result obtained by using the standard prescription of adiabatic subtraction, i.e. by removing terms up to the second adiabatic order. The dot dashed line represents the renormalized result obtained with the method of the optimal truncation, i.e. removing, for superhorizon modes, only the zeroth order of the WKB series.}
	\label{fig:massivedesitter}
\end{figure}

In Figure~\ref{fig:massivedesitter}  we show the non-renormalized power spectrum and the two versions of the normalized one obtained by using the two different prescritions. As we can see the use of the prescription~\cite{Dabrowski:2014ica,Dabrowski:2016tsx} eliminates the first of the ``undesirable features'' listed in the Introduction, namely the fact that the standard prescription for adiabatic subtraction leads to a significant running of the renormalized power spectrum for $m\ll k/a\lesssim H$. With the method of optimal truncation on the other hand we get an almost constant spectrum.

\subsection{A massless scalar in inflation, followed by reheating}

We have shown in Section~\ref{subsec:massless} above that, by applying the prescription of~~\cite{Birrell:1982ix,Parker:2009uva}, the power spectrum of a massless scalar in exact $3+1$-dimensional de Sitter space is identically vanishing. It is however easy to see that super-horizon modes for which $k\ll a\, H$ are not evolving adiabatically in this system, since 
\begin{align}
\frac{1}{\omega^2}\left|\frac{d\omega}{d\theta}\right|=2\,\frac{aH}{k}\gg 1\,,\qquad \qquad {\rm for}\,k\ll a\, H\,.
\end{align}

It is more interesting to consider the realistic situation where slow roll inflation is followed by reheating and a radiation dominated Universe.

To study such a system, we consider a massless scalar field on the top of an FLRW Universe whose scale factor evolution is given by
\begin{align}\label{eq:aoftcompl}
a(t)=2\,e^{Ht}\frac{(1+H^2t^2)^{1/4}}{e^{Ht}+2\,(1+H^2t^2)^{1/4}}\,,
\end{align} 
which gives a de Sitter metric $a(t)\sim e^{Ht}$ for $t\ll -1/H$ and radiation dominated cosmology $a(t)\sim 2\sqrt{Ht}$ for $t\gg 1/H$. This choice of the form of $a(t)$ also  leads to $\dot{H}<0$ at all times, as required by energy conditions.

For a scale factor given by eq.~(\ref{eq:aoftcompl}), the proper frequency $\omega_k(\theta)=k\,a^2(\theta)$ has complex zeros at $\theta=\theta_c$ and $\theta=\theta_c^*$, corresponding to $Ht=\pm i$, and we get
\begin{align}
F_k^{(0)}=-i\int_{\theta_c}^{\theta_c^*}\omega_k(\theta)\,d\theta=-i\,k\int_{-i}^{i}a(t)^2\,\frac{dt}{a(t)^3}\simeq 2.88\,\frac{k}{H}\,.
\end{align}

Numerical evaluation shows that the Stokes line from $\theta_c$ to $\theta_c^*$ crosses the real axis at a value of $\theta$ corresponding to $t\simeq .34\,H^{-1}$, which can be identified as the time at which production of quanta of $\phi$ occurs.

We have computed numerically the spectrum of $\phi$ as a function of time. The bare spectrum is shown, for $t=0$ and $t=100\,H^{-1}$ in Figure~\ref{fig:baresp}. Note that, as expected, the power spectrum converges to $(H/2\pi)^2$ in the limit of long wavelengths, while it goes as $k^2/a^2$ at short wavelengths.

In Figure~\ref{fig:subsp} we show, instead, the power spectrum for the field $\phi$ obtained by subtracting a regularized optimally truncated version of $W(t)$. 

The regularization is built as follows. We define the functions $\tilde{W}_{(2i)}(t)$ through
\begin{align}
W_{(2n)}(t)=k\left[\tilde{W}_{(0)}(t)+\frac{\tilde{W}_{(2)}(t)}{k^2}+\frac{\tilde{W}_{(4)}(t)}{k^4}+\ldots+\frac{\tilde{W}_{(2n)}(t)}{k^{2n}}\right]\,,
\end{align}
where $\tilde{W}_{(0)}(t)=a(t)^2$. Next, we define weighted averages of $|\tilde{W}_{(2i)}(t)|$ as
\begin{align}
\hat{W}_{(2i)}(t)=\int_{-\infty}^\infty|\tilde{W}_{(2i)}(t+t_1)|\,e^{-H^2t_1^2}\,dt_1\,,
\end{align}
and the regularized Heaviside $\Theta$ functions as
\begin{align}
\Theta_Q(t;\,2m,\,2n)=\frac12\left[1+\tanh\left(2Q\frac{\hat{W}_{(2m)}(t)-\hat{W}_{(2n)}(t)}{\hat{W}_{(2m)}(t)+\hat{W}_{(2n)}(t)}\right)\right]
\end{align}
that converge to the Heaviside step function for $Q\to\infty$. In our numerical evaluation we set $Q=5$.

Finally, the regularized optimal truncation of $W(t)$ is obtained as
\begin{align}\label{eq:w_opt_reg}
W^{\rm optimal}_{\rm reg}(t)&=k\,\tilde{W}_{(0)}(t)\,\Theta_5(t;\,2,\,0)+k\,\left(\tilde{W}_{(0)}(t)+\frac{\tilde{W}_{(2)}(t)}{k^2}\right)\,\Theta_5(t;\,0,\,2)\,\Theta_5(t;\,4,\,2)+\nonumber\\
&+k\left(\tilde{W}_{(0)}(t)+\frac{\tilde{W}_{(2)}(t)}{k^2}+\frac{\tilde{W}_{(4)}(t)}{k^4}\right)\,\Theta_5(t;\,0,\,2)\,\Theta_5(t;\,2,\,4)\,\Theta_5(t;\,6,\,4)+\ldots
\end{align}
so that $W^{\rm optimal}_{\rm reg}(t)=k\,\tilde{W}_{(0)}(t)$ for $\hat{W}_{(2)}(t)\gg \hat{W}_{(0)}(t)$, and $W^{\rm optimal}_{\rm reg}(t)=k\left(\tilde{W}_{(0)}(t)+\frac{\tilde{W}_{(2)}(t)}{k^2}\right)$ for $\hat{W}_{(0)}(t)\gg \hat{W}_{(2)}(t)$ and $\hat{W}_{(2)}(t)\ll \hat{W}_{(4)}(t)$ (i.e., $\hat{W}_{(2)}(t)$ is a local minimum of the $\hat{W}_{(2i)}(t)$), etc...

The weighted average $\hat{W}_{(2i)}(t)$ is introduced to eliminate spurious effects originating from the fact that the functions  $\tilde{W}_{(2i)}(t)$ are generally oscillating, and therefore cross zero and appear to be small even if they have a large amplitude, for the relevant values of time $t$. The regularized $\Theta$ function is used to lead to a smooth spectrum. 

As Figure~\ref{fig:subsp} shows, the use of this regularized optimal truncation eliminates the  singularity in the power spectrum around $k\simeq .5\,H$ that emerges when one uses  the second order adiabatic subtraction, which originates from the fact that for that value of $k$ one has $W_{(2)}(k,\,t=0)\simeq 0$.

On the other hand, the expression for $W^{\rm optimal}_{\rm reg}(t)$ rapidly converges to $W_{(2)}(k,\,t)$ as $t$ grows. In Figure~\ref{fig:subspfinal} we show the subtracted spectra for $t=100\,H^{-1}$. However, already at $t=5\,H^{-1}$ the two subtracted spectra are indistinguishable. 

\begin{figure}[t]
\centering
\includegraphics[scale=.9]{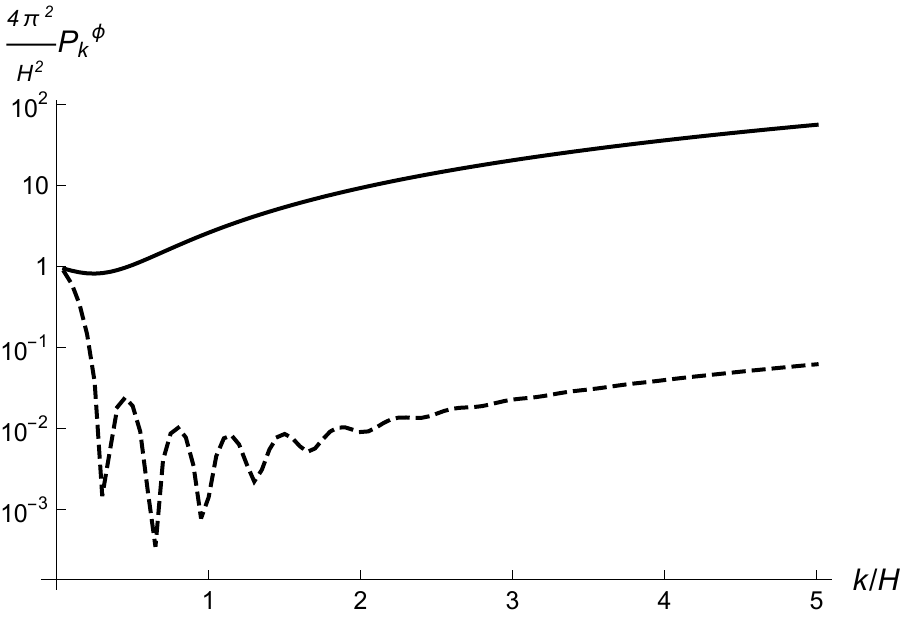} 
\caption{The unsubtracted power spectrum (in units of $(H/2\pi)^2$) of a massless scalar in an expanding Universe with the expression of the scale factor given by eq.~(\ref{eq:aoftcompl}). The spectrum is evaluated at $t=0$ (solid line) and at $t=100\,H^{-1}$ (dashed). Modes with $k\gtrsim .9\,H$ satisfy $k/a>\dot{a}/a$ for the entire evolution, so that they never cross the horizon. For those modes the power spectrum goes as $k^2/a^2$.}
\label{fig:baresp}
\end{figure}

\begin{figure}[t]
\centering
\includegraphics[scale=.9]{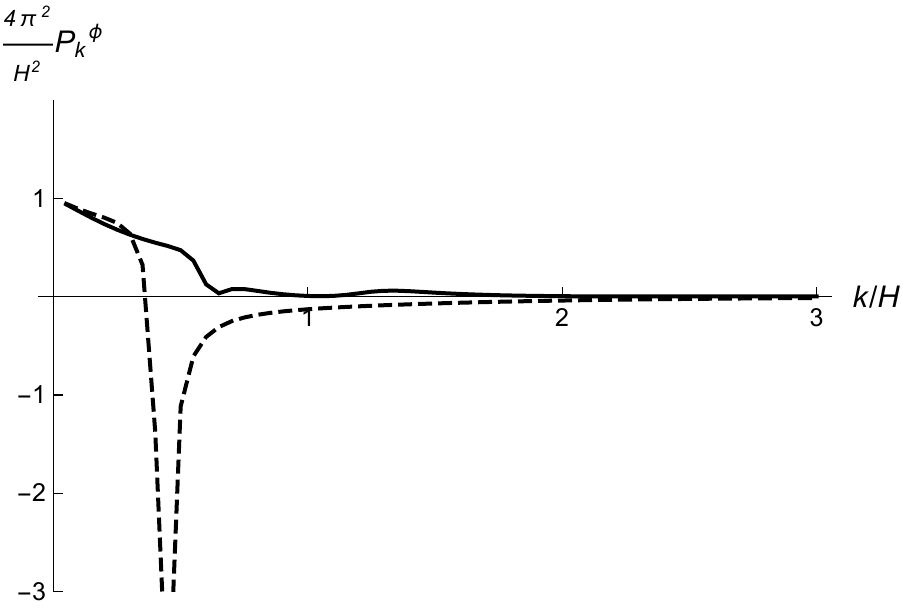} 
\caption{The subtracted power spectra (in units of $(H/2\pi)^2$) of a massless scalar in an expanding Universe with the expression of the scale factor given by eq.~(\ref{eq:aoftcompl}). The spectra is evaluated at $t=0$. Solid: the spectrum obtained by subtracting a regularized version (see the main text leading to eq.~(\ref{eq:w_opt_reg})) of the optimally truncated expression for $W(t)$. Dashed: the spectrum obtained by subtracting $W_{(2)}(t)$.}
\label{fig:subsp}
\end{figure}

\begin{figure}[t]
\centering
\includegraphics[scale=1]{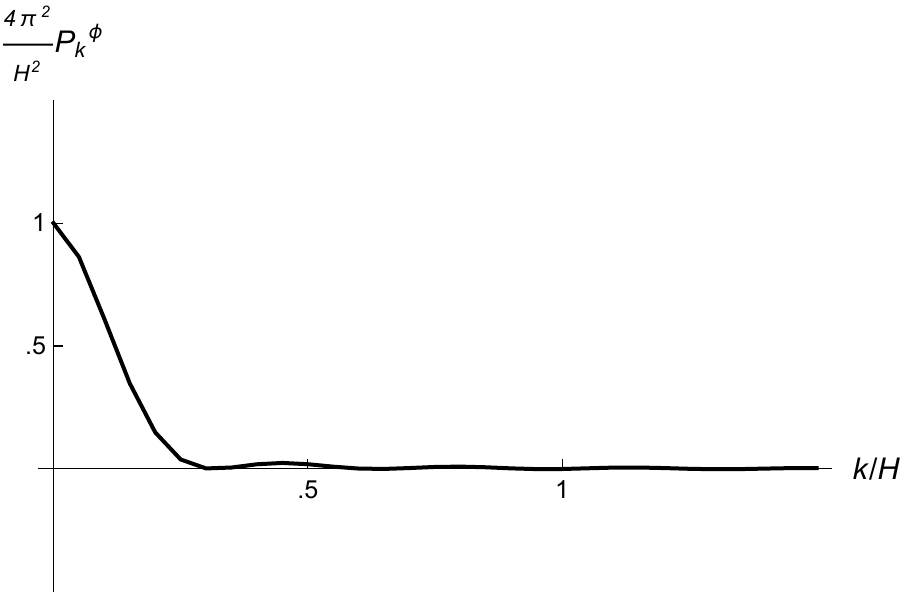} 
\caption{The subtracted power spectra (in units of $(H/2\pi)^2$) of a massless scalar in an expanding Universe with the expression of the scale factor given by eq.~(\ref{eq:aoftcompl}). The spectrum is evaluated at $t=100\,H^{-1}$. The spectrum obtained by subtracting $W^{\rm optimal}_{\rm reg}(t)$ and that obtained by subtracting $W_{(2)}(t)$ are, for this large value of $t$, numerically equivalent.}
\label{fig:subspfinal}
\end{figure}

\section{Conclusions and discussion}%
\label{sec:conclusions}

In this work we have revisited the method of adiabatic subtraction for the renormalization of the power spectrum of scalar perturbations generated during inflation. First, we have reviewed the equivalence between adiabatic subtraction and  normal ordering of the time-dependent creation/annihilation operators that instantaneously diagonalize the Hamiltonian. This requires the proper frequency of the mode functions to be real at all times, and we have found a definition of time that guarantees this condition to be satisfied. 

The main question we have tackled is to what order one should truncate the WKB expansion of the adiabatic modes that have to be subtracted to provide the renormalized result. The standard prescription~\cite{Parker:1969au,Parker:1971pt,Parker:1974qw,Birrell:1982ix,Parker:2009uva} to truncate the WKB series to the lowest order that allows to cancel all divergences, while having the advantage of leading to (relatively) simple calculations, can generate artifacts at intermediate momenta~\cite{Parker:2007ni}. An alternative option~\cite{Dabrowski:2014ica,Dabrowski:2016tsx} is based on the fact that the WKB approximation is generally an asymptotic one, which naturally  results into truncating the WKB series to the value that gives the closest approximation to the ``actual expression'' of the Bogolyubov coefficient $\beta$, that has been argued to take a universal form~\cite{dingle,Barry:1989zz,Barry:1990}.

We applied these prescriptions to the massless and massive minimally coupled scalar field on exact de Sitter space in flat slicing, and we have compared the resulting power spectra. In the massless case, the fact the WKB series can be resummed implies that we do not have particle production, and the renormalized power spectra turn out to vanish using both prescriptions. This result, in line with those in~\cite{Habib:1999cs,Parker:2009uva} is due to the fact that de Sitter space in flat slicing, unlike de Sitter space in closed slicing, does not have a built-in infrared cutoff, which is responsible for the growth $\langle\phi(\bx,\,t)^2\rangle\propto H^3\,t$.

In the massive case the results are quite different: optimal truncation requires, for super-horizon modes $k\lesssim a\,H$, to remove only the zeroth order WKB contribution, leading to the standard quasi scale-invariant spectrum, while the usual adiabatic renormalization removes terms up to the second adiabatic order and leads to a significant running for scales $m\ll k/a\lesssim H$. Subsequently,  we applied our method to a more complicated system, a massless scalar field which undergoes a phase of slow roll inflation followed by reheating and a subsequent radiation dominated era. In this case, even though the field is massless, the power spectrum is non vanishing and takes the standard expression $\sim (H/2\pi)^2$ at large scales, while at shorter scales the prescription based on the optimal truncation of the WKB series eliminates some of the artifacts that are generated by the standard second order truncation.

The prescription to truncate the WKB expansion at its optimal order is justified by the consideration that the process of particle production can be identified with the Stokes phenomenon. The connection formula that extends the WKB approximation to the whole complex plane is referring to asymptotic series truncated at their least terms. More recently, it has been found that truncating the WKB series at the optimal order, {\em as long as such an optimal order is much larger than $O(1)$}, leads to a sum that approximates well the universal behavior across the Stokes line found by Dingle and Berry~\cite{dingle,Barry:1989zz,Barry:1990}. This theory is effective when the regions of non-adiabaticity in the complex plane are small enough.  In this work, we used the optimal truncation of the WKB series even when it happens at a low order, i.e., when the thickness of the Stokes lines becomes comparable with their length or when two regions of non-adiabaticity overlap.  It would be interesting to see whether this choice is justified by an argument analogous to that of~\cite{dingle,Barry:1989zz,Barry:1990}. 

\acknowledgments We thank Paolo Creminelli, John Donoghue, David Kastor, Mehrdad Mirbabayi, Marco Peloso, Borna Salehian, Luca Santoni, Gianmassimo Tasinato and Jennie Traschen for useful discussions. A special thanks goes to Gerald Dunne for his patient explanations. This work is partially supported by the US-NSF grants PHY-1820675 and PHY-2112800.


\end{document}